\documentclass[a4paper,10pt]{article} 
\usepackage[utf8]{inputenc}
\usepackage{authblk}
\usepackage{fullpage}
\usepackage{graphicx}
\usepackage{xcolor}
\usepackage{algorithm}
\usepackage{algpseudocode}
\usepackage{amssymb}
\usepackage{amsthm}
\usepackage{amsmath}
\usepackage{url}
\usepackage{bbold}
\usepackage[square,numbers,comma,compress]{natbib} 
\usepackage[colorlinks,urlcolor=red,citecolor=red,linkcolor=blue,pdftex, pdfauthor = Lukas Exl]{hyperref}
\usepackage{tabularx}
\usepackage{setspace} 
\usepackage{listings}
\lstdefinestyle{customc}{
  belowcaptionskip=1\baselineskip,
  breaklines=true,
  frame=L,
  xleftmargin=\parindent,
  language=MATLAB,
  showstringspaces=false,
  basicstyle=\footnotesize\ttfamily,
  keywordstyle=\bfseries\color{green!40!black},
  commentstyle=\itshape\color{purple!40!black},
  identifierstyle=\color{blue},
  stringstyle=\color{orange},
}

\lstdefinestyle{customasm}{
  belowcaptionskip=1\baselineskip,
  frame=L,
  xleftmargin=\parindent,
  language=[x86masm]Assembler,
  basicstyle=\footnotesize\ttfamily,
  commentstyle=\itshape\color{purple!40!black},
}
\title{\Large\textbf{The extrapolated explicit midpoint scheme for variable order and step size controlled integration of the Landau-Lifschitz-Gilbert equation}}
\author[1,2]{Lukas Exl \thanks{\texttt{lukas.exl@univie.ac.at}}}
\author[3]{Norbert J. Mauser} 
\author[4]{Thomas Schrefl}
\author[5]{Dieter Suess}
\affil[1]{\small Fak. Mathematik, Univ. Wien, 1090 Vienna, Austria.} 
\affil[2]{Inst. of Solid State Physics, TU Wien, 1040 Vienna, Austria.} 
\affil[3]{Wolfgang Pauli Institute c/o Fak. Mathematik, Univ. Wien, 1090 Vienna, Austria.}
\affil[4]{Center of Integrated Sensor Systems, Danube Univ. Krems, 2700 Wr. Neustadt, Austria}
\affil[5]{CD-Laboratory for Advanced Magnetic Sensing and Materials, TU Wien, 1040 Vienna, Austria}
\begin{document}
\date{}
\maketitle
\noindent\textbf{Abstract.} A practical and efficient scheme for the higher order integration of the Landau-Lifschitz-Gilbert (LLG) equation is presented. The method is based on extrapolation of the two-step explicit midpoint rule and 
incorporates adaptive time step and order selection. We make use of a piecewise time-linear stray field approximation to reduce the necessary work per time step. The approximation to the interpolated operator is embedded into the 
extrapolation process to keep in step with the hierarchic order structure of the scheme. We verify the approach by means of numerical experiments on a standardized NIST problem and compare with 
a higher order embedded Runge-Kutta formula. The efficiency of the presented approach increases when the stray field computation takes a larger portion of the costs for the effective field evaluation.\\

\noindent\textit{Keywords}: Landau-Lifschitz-Gilbert equation, extrapolation method, explicit midpoint scheme, variable\\ order method, micromagnetics 
\section{Introduction}
Micromagnetics is a continuum theory of ferromagnetic materials located between classical Maxwell's theory of electromagnetism and quantum theory \cite{brown1963micromagnetics}. 
A ferromagnetic system is described by the total magnetic energy of its magnetic distribution, which is modeled as a continuous vector field within the magnetic material. Typical length scales, that can be resolved 
by micromagnetic models, are in the range of a few nanometers to micrometers, which is too large for atomistic spin dynamics. 
On the other hand, these length scales are large enough for computer simulations of magnetic data storage systems like hard discs \cite{suess2015fundamental,kovacs2016numerical} or random access memory \cite{makarov2012fast} and 
high performance permanent magnets \cite{sepehri2013high,bance2014high}.\\
The fundamental equation for dynamic processes of the magnetization in a magnetic body $\Omega \subset \mathbb{R}^3$, a vector field $\boldsymbol{M}(x,t) = M_s \boldsymbol{m}(x,t),\,|\boldsymbol{m}(x,t)| = 1$ 
depending on the position $x \in \Omega$ and time $t\in \mathbb{R}$, is the Landau-Lifschitz-Gilbert equation. It is given in explicit form as \cite{kronmueller} 
\begin{align}\label{LLGM}
 \frac{\partial \boldsymbol{M}}{\partial t} &\, = -\frac{\gamma_0}{1+\alpha^2} \, \boldsymbol{M} \times \boldsymbol{H} - \frac{\alpha\,\gamma_0}{(1+\alpha^2) M_s} \, \boldsymbol{M} \times \big(\boldsymbol{M} \times \boldsymbol{H}\big),
\end{align}
where $\gamma_0$ is the gyromagnetic ratio, $\alpha$ the damping constant and $\boldsymbol{H}$ the effective field, which is the sum of nonlocal and local fields such as the stray field and the exchange field, respectively.\\
Typically, equation \eqref{LLGM} is numerically treated by a spatial semi-discrete approach \cite{d2005geometrical,suess2002time,donahue1999oommf}. 
We mention here, that in recent years also lower order finite element methods for the LLG equation were developed along with convergence analysis of weak solutions \cite{alouges2006convergence,bartels2006convergence,kritsikis2014beyond}. 
The computational main difficulty for numerics of the LLG equation arises from the expensive right hand side evaluation, mostly 
due to the nonlocal part in the effective field, namely the stray field. Several numerical methods were developed for the stray field calculation \cite{strayfield_review}. They either rely on a scalar potential or a field-based approach and scale at best 
linearly with the number of discrete magnetic spins or computational units. Nevertheless, the amount of computational costs for this calculation is typically $80-90 \%$ of that of the (total) effective field. 
Hence, it is desirable to develop numerical schemes for micromagnetics that try to avoid excessive field evaluations, while also maintaining accuracy and efficiency. 
In the large $\alpha$ case, equation \eqref{LLGM} degenerates to a steepest descent method for minimizing the total energy, owing to the Lyapunov structure of the LLG equation \cite{d2005geometrical}. 
In this case, steepest descent methods \cite{exl2014labonte} and conjugate gradient variants \cite{fischbacher2017conjugate} were recently developed, which already require fairly optimal amounts of field evaluations.  
In moderately damped cases high accuracy and large time steps can be achieved by either higher order non-stiff integrators, 
as Runge-Kutta methods \cite{donahue1999oommf} or implicit schemes such as the midpoint method \cite{d2005geometrical} or backward differentiation formulas (BDF) \cite{suess2002time}. 
Other methods, as semi-analytic and geometric integration and projected Gauss-Seidel, can be found in the review \cite{garcia2007numerical} and references therein.   
Implicit schemes require special treatment of the (non)linear systems of equations, 
which have to be solved each time step. These systems are basically dense owing to the nonlocal stray field. 
Efficiency will also strongly rely on the successful application of preconditioners, which might also have to be recomputed during integration \cite{suess2002time}. Typical time step lengths reached by implicit second order methods are in the 
range of picoseconds, while those of explicit higher order schemes lie in the range of several femtoseconds. Hence, implicit schemes will need fewer derivative evaluations for establishing the time step equations, 
but shift the computational task to the numerical treatment of the (non)linear systems. These systems should be solved accurately and efficient and might require additional field evaluations as well. 
On the other hand, higher order explicit schemes require larger amounts of field evaluations per step, 
for instance, the classical $4$th order Runge-Kutta scheme with $5$th order local error estimate requires $6$ evaluations per time step and 
an $8$th order Dormand-Prince formula with $7$th order local error estimate requires already $13$ evaluations per time step \cite{hairer_old}. 
These explicit methods also incorporate adaptive step size selection, which provides them with additional efficiency and robustness.  
Equation \eqref{LLGM} is non-stiff for largely homogeneous materials and simple geometries \cite{tsiantos2001stiffness}, but might only get stiffer if grain structures are also modeled \cite{suess2002time}. 
For instance, OOMMF \cite{donahue1999oommf}, likely the most widely used micromagnetic simulation package, uses explicit (non-stiff) embedded Runge-Kutta formulas of different selectable order for the integration routines of the spatially semi-discretized equation \eqref{LLGM}. 
We will construct an explicit higher order scheme for \eqref{LLGM} that is especially cheap in terms of stray field evaluations, while maintaining higher order properties for iterates and local error estimates.
This is achieved by exploiting extrapolation for the Gragg method \cite{gragg}, also known as explicit midpoint scheme. The meta-principle of 
(Richardson) extrapolation applies to computed quantities, which depend on a parameter like  a mesh or step size. 
Consider, for instance, a spatially semi-discretized version of \eqref{LLGM} and a prescribed initial magnetization. Now, consider the error $\boldsymbol{e}(t;h)$ of a numerical approximation $\boldsymbol{\eta}(t;h)$ of the magnetization 
$\boldsymbol{M}(t)$ at some time $t$ obtained from an iteration scheme (some ODE solver) that uses a step size $h$. If the error possess an asymptotic expansion in $h$ 
\begin{align}
 \boldsymbol{e}(t;h) := \boldsymbol{\eta}(t;h) - \boldsymbol{M}(t) = \boldsymbol{c}_1(t)\, h^{\beta p} + \boldsymbol{c}_2(t)\, h^{\beta (p+1)} + \mathcal{O}(h^{\beta(p+2)}),
\end{align}
we could recompute the approximation with reduced step size, e.g., halved $h/2$, and establish a new extrapolated approximation according to
\begin{align}\label{extrapsol}
 \boldsymbol{\eta}(t;h,h/2) := \boldsymbol{\eta}(t;h/2) + \frac{\boldsymbol{\eta}(t;h/2) - \boldsymbol{\eta}(t;h)}{2^{\beta p} -1 }.
\end{align}
For the new approximation the lowest error term is canceled, that is 
\begin{align}
  \boldsymbol{\eta}(t;h,h/2) = \boldsymbol{M}(t) + \mathcal{O}(h^{\beta (p+1)}).
\end{align}
This is especially efficient if $\beta>1$, which is true, with $\beta = 2$, for symmetric methods \cite{stetter1970symmetric, deuflhard1985recent}. Natural candidates are the midpoint scheme or 
the trapezoidal rule, which are both implicit and second order in time. 
Due to the implicit nature, the error expansion of such methods only holds within the numerical accuracy of the solutions of the (non)linear systems. On the other hand, the Gragg method is a symmetric  explicit two-step scheme, 
which is therefore ideal for establishing an exact extrapolation approach for the LLG equation \eqref{LLGM}. This is done in a triangular Aitken-Neville scheme for polynomial extrapolation, 
which offers a natural way for adaptive step size and order selection via computationally available local error estimates and the hierarchic order structure. 
The well-known Gragg-Bulirsch-Stoer (GBS) algorithm \cite{bulirsch1966numerical} for general non-stiff initial value problems is based on the Gragg method and rational function extrapolation. However, it turned out that polynomial extrapolation 
is almost always more effective \cite{hairer}. While extrapolation methods for initial value problems are designed for highly accurate nuemrical solutions, the drawback is the increased amount of derivative evaluations 
because of successive step doubling. In this paper we construct higher order schemes for \eqref{LLGM} via polynomial extrapolation of the Gragg method and save expensive stray field evaluations, 
while simultaneously maintaining the order properties for the iterates and the local error estimates. 
This is achieved by treating a version of equation \eqref{LLGM} with time-linear stray field, where the computational realization of the linear interpolation is incorporated in the extrapolation procedure. 
We combine the resulting hierarchic structure of higher order schemes in an interplaying step size and order adaptive procedure.\\   
In the following two sections we will clarify the problem setting and give details to the extrapolated Gragg method. Section~\ref{taming} explains the approach for taming the complexity of the extrapolation scheme. A further section 
is dedicated to the adaptive step size and order selection. Finally, we validate the method in terms of accuracy and efficiency on variations of 
the NIST $\mu$MAG Standard problem $\#4$ \cite{mumag4} and also compare it to a higher order Dormand-Prince formula.
\section{Problem setting}
Let $\Omega \subset \mathbb{R}^3$ denote a magnet and $\boldsymbol{m}: \, \Omega \rightarrow \mathbb{R}^3$ the reduced (dimensionless) magnetization. 
The \textit{magnetic Gibbs free energy} (in dimensionless form) is  given by \cite{kronmueller}
\begin{equation}\label{gibbs_energy}
\begin{aligned}
 e_{tot}(\boldsymbol{m})
  =\frac{1}{|\Omega|}\Big( \frac{A}{\mu_0 M_s^2}\int_{\Omega} |\nabla \boldsymbol{m}|^2 \,\text{d}x - \frac{1}{2} \int_{\Omega} \boldsymbol{m} \cdot \boldsymbol{h}_s (\boldsymbol{m}) \, \text{d}x - 
  \frac{K_1}{\mu_0 M_s^2} \int_{\Omega} (\boldsymbol{a} \cdot \boldsymbol{m})^2 \, \text{d}x - \int_{\Omega} \boldsymbol{m} \cdot \boldsymbol{h}_{ext} \, \, \text{d}x \Big),
\end{aligned}
\end{equation}
that is the sum of exchange-, demagnetizing-, (uniaxial/first order) anisotropy- and external energy, respectively. 
Here $\mu_0$ is the vacuum permeability, $M_s$ the saturation magnetization, $A$ the exchange constant, $K_1$ the first magnetocrystalline anisotropy constant and $\boldsymbol{a}$ the unit vector parallel to the easy axis. 
Further, $\boldsymbol{h}_s$ is the (dimensionless) stray field, which is defined by the magnetostatic Maxwell equation $-\nabla \cdot \boldsymbol{h}_s = \nabla \cdot \boldsymbol{m} \,\,\text{in}\,\, \mathbb{R}^3$ and $\boldsymbol{h}_{ext} := \boldsymbol{H}_{ext}/M_s $ is the (dimensionless) external field.\\
The Landau-Lifschitz-Gilbert (LLG) equation \cite{kronmueller,brown1963micromagnetics,aharoni2000introduction} describes the time evolution of the magnetization and is given in a dimensionless and explicit form as
\begin{align}\label{LLG}
 \frac{\partial \boldsymbol{m}}{\partial\tau} = -\frac{1}{1+\alpha^2} \, \boldsymbol{m} \times \boldsymbol{h}(\boldsymbol{m}) - \frac{\alpha}{1+\alpha^2} \, \boldsymbol{m} \times \big(\boldsymbol{m} \times \boldsymbol{h}(\boldsymbol{m})\big),
\end{align}
where $\alpha>0$ is the (dimensionless) damping constant and $\boldsymbol{m} = \boldsymbol{m}(x,t): \, \Omega \times [0,T] \rightarrow \mathbb{R}^3$ the time-dependent magnetization.  
The parameter $\tau$ in equation \eqref{LLG} is dimensionless owing to the relation $\tau = M_s \gamma_0 \, t$ to the physical time $t$, where $\gamma_0$ is the gyromagnetic ratio. 
The effective field $\boldsymbol{h}$ is defined via the functional derivative of the energy
\begin{align}\label{heff}
\boldsymbol{h}(\boldsymbol{m}) := -|\Omega|\,\frac{\delta e_{tot}}{\delta m} = \frac{2A}{\mu_0 M_s^2} \, \Delta \boldsymbol{m} + \boldsymbol{h}_s(\boldsymbol{m}) + \frac{2 K_1}{\mu_0 M_s^2} \, (\boldsymbol{a}\cdot \boldsymbol{m})\,\boldsymbol{a} + \boldsymbol{h}_{ext}.
\end{align}
Equation \eqref{LLG} is supplemented with the 
initial condition 
\begin{align}
 \boldsymbol{m}(x,0) = \boldsymbol{m}^{(0)}(x)
\end{align}
and the boundary condition 
\begin{align}\label{bcs}
 \frac{\partial \boldsymbol{m}}{\partial \boldsymbol{\nu}} = 0, \quad x \in \partial{\Omega},
\end{align}
where $\boldsymbol{\nu}$ is the outward unit normal on the boundary $\partial{\Omega}$. The LLG equation preserves the magnitude of the initial magnetization as can be seen by scalar multiplication with $\boldsymbol{m}$.\\
We treat the spatially semi-discretized LLG equation in the spirit of ordinary differential equations as in \cite{d2005geometrical}, where $\boldsymbol{m}$ is represented by a discrete mesh vector in each of the $N$ computational units, 
e.g., cubical/rectangular computational cells. Also, a linear finite element approach on tetrahedral meshes can lead to a similar system of ordinary differential equations \cite{suess2002time}. The discrete mesh vectors are collected in a long vector of size $3N$, as well as the discrete effective field components evaluated at the discrete magnetization. 
The exchange field is discretized by symmetric second order finite differences where the boundary condition \eqref{bcs} is taken into account. The stray field is nonlocal and computed by the algorithm described in \cite{exl2012fast}, 
which is based on a scalar potential and accelerated by FFT as introduced in \cite{exl2014fft}. The other components in \eqref{heff} permit evident discrete local representations. 
In this sense, we treat equation \eqref{LLG} as a system of ordinary differential equations for the $3N$ components of the long discrete magnetization vector
\begin{eqnarray}\label{LLGd}
\begin{aligned}
 \frac{\partial \boldsymbol{m}_j}{\partial\tau} &\, = -\frac{1}{1+\alpha^2} \, \boldsymbol{m}_j \times \boldsymbol{h}_j(\boldsymbol{m}) - \frac{\alpha}{1+\alpha^2} \, \boldsymbol{m}_j \times \big(\boldsymbol{m}_j \times \boldsymbol{h}_j(\boldsymbol{m})\big), \quad j = 1,\hdots,N,\\
 \boldsymbol{m}(0) &\,= \boldsymbol{m}^{(0)},
 \end{aligned}
\end{eqnarray}
where $\boldsymbol{m} = (\boldsymbol{m}_1^T,\boldsymbol{m}_2^T,\hdots,\boldsymbol{m}_N^T)^T \in \mathbb{R}^{3N}$.\\

Numerical algorithms for the LLG equation, which do not preserve the unit norm constraint of the magnetization inherently, consider renormalization of the discrete magnetic spins after each iteration or if some accuracy tolerance is violated.  
We present our method without renormalization, hence the deviation from the unit norm constraint may also serve as a measure of accuracy. However, there is no limitation to it in the forthcoming method, so that renormalization could be 
incorporated. 

\section{The extrapolated Gragg method}\label{method}

\begin{figure}[hbtp]
\centering 
\includegraphics[scale=0.7]{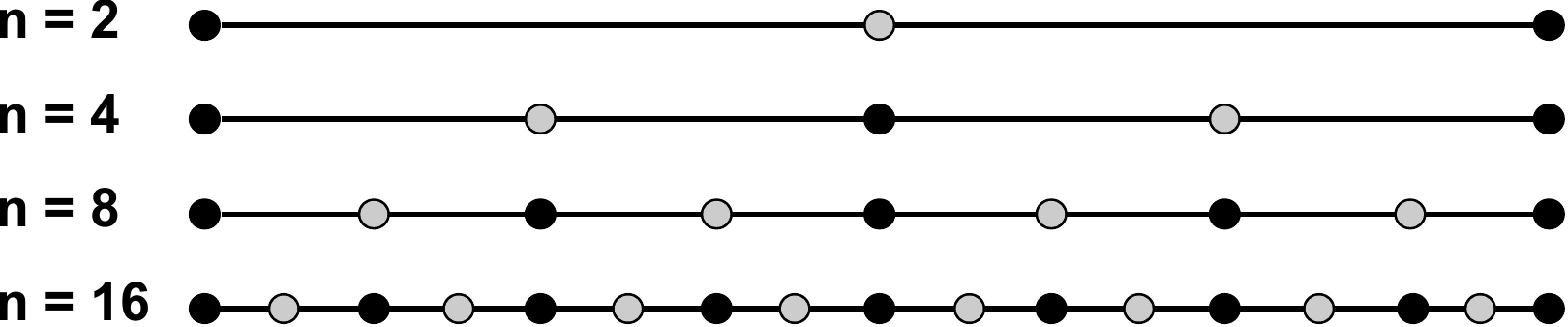}    
\caption{First four extrapolation levels using the Romberg sequence. Black dots correspond to even indices and gray dots to odd indices.}\label{fig:romberg}
\end{figure}

Our scheme is based on the explicit two-step midpoint rule (Gragg method) \cite{gragg,hairer} for the initial value problem \eqref{LLGd} given in the abbreviated form
\begin{eqnarray}\label{ivp}
\begin{aligned}
\boldsymbol{m}^\prime(t) &\, = F\big(\boldsymbol{m}(t)\big),\\
\boldsymbol{m}(t_0) &\,= \boldsymbol{m}^{(0)}. 
\end{aligned}
\end{eqnarray}
Let the desired approximation to \eqref{ivp} at $t = t_0 + H,\,H > 0$ be denoted with $\boldsymbol{m}_h(t)$ where $h:= H/n$ and $n$ an even number. Gragg's midpoint rule reads  
\begin{eqnarray}\label{Gragg}
\begin{aligned}
 \boldsymbol{m}^{(1)} &\,= \boldsymbol{m}^{(0)} + h \,F(\boldsymbol{m}^{(0)})\\
 \boldsymbol{m}^{(\nu+1)} &\,= \boldsymbol{m}^{(\nu-1)} + 2h\,F(\boldsymbol{m}^{(\nu)})\quad \nu = 1,2,\hdots,n\\
 \boldsymbol{m}_h(t) &\, = \frac{1}{4} (\boldsymbol{m}^{(n-1)} + 2\boldsymbol{m}^{(n)} + \boldsymbol{m}^{(n+1)}) = \frac{1}{2}\big(\boldsymbol{m}^{(n)} + \boldsymbol{m}^{(n-1)} + h\, F(\boldsymbol{m}^{(n)})\big).
\end{aligned}
\end{eqnarray}
The method is consistent of order $2$ and the equivalent one-step scheme is symmetric \cite{stetter1970symmetric}. It therefore possesses an asymptotic error expansion in even powers of $h$, provided the function $F(\boldsymbol{m})$ is sufficiently smooth, that is 
\begin{eqnarray}\label{h2expansion}
\begin{aligned}
 \boldsymbol{m}(t_0+\nu h) - \boldsymbol{m}_h(t_0+ \nu h) &\,= \sum_{j=1}^\ell \boldsymbol{a}_j(t_0+\nu h)\,h^{2j} + h^{2j+2} \boldsymbol{c}(t_0+\nu h;h), \quad \nu \,\,\text{even} \\
 \boldsymbol{m}(t_0+\nu h) - \boldsymbol{m}_h(t_0+ \nu h) &\,= \sum_{j=1}^\ell \boldsymbol{b}_j(t_0+\nu h)\,h^{2j} + h^{2j+2} \widetilde{\boldsymbol{c}}(t_0+\nu h;h), \quad \nu \,\,\text{odd},
\end{aligned}
\end{eqnarray}
where the expansions are different for even and odd indices $\nu$. There holds $\boldsymbol{a}_j(t_0) = 0$ in the even case, but $\boldsymbol{b}_j(t_0) \neq 0$ in the odd case. 
The existence of an error expansion in powers of $h^2$ is crucial for the efficiency of the Richardson extrapolation based method in the forthcoming. Due to Gragg \cite{gragg} the (first order) explicit Euler starting step 
is enough for guaranteeing the $h^2$-expansion. 
The averaging in the last line of \eqref{Gragg} is a smoothing step, which should originally reduce 'weak stability' by eliminating the lowest corresponding error term. 
It is actually not needed for that purpose, if the Gragg method is applied together with extrapolation, which cancels theses error terms anyway. 
Omitting the smoothing would save one $F$-evaluation and also maintains the asymptotic expansions, where in this case we would simply have $\boldsymbol{m}_h(t) := \boldsymbol{m}^{(n)}$. In the algorithm for the LLG equation 
we will use the smoothing step at no additional cost, since the $F$-evaluation at the interval end is provided for different reason. 
The well known Gragg-Bulirsch-Stoer (GBS) algorithm \cite{bulirsch1966numerical} for general non-stiff initial value problems \eqref{ivp} is based on Gragg's midpoint scheme and rational extrapolation. 
However, it turned out that polynomial extrapolation is more efficient \cite{hairer}. The extrapolated Gragg method is built up of approximations $\boldsymbol{\mu}_{\ell,1} := \boldsymbol{m}_{h_\ell}(t),\,\,\ell=1,\,2,\hdots$, where $h_\ell = H/n_\ell$ and 
$n_\ell$ an increasing sequence of even numbers, 
e.g. the Romberg (power-two) sequence $\{2,4,8,16,32,64,\hdots$\}, see Fig.~\ref{fig:romberg}. 
Owing to the $h^2$-expansion \eqref{h2expansion} the Aitken-Neville algorithm on level $\ell$ leads to 
\begin{align}\label{AitkenNeville}
 \boldsymbol{\mu}_{\ell,m} = \boldsymbol{\mu}_{\ell,m-1} + \frac{\boldsymbol{\mu}_{\ell,m-1} - \boldsymbol{\mu}_{\ell-1,m-1}}{(n_\ell/n_{\ell-m+1})^2-1}\quad m=1,\hdots,\ell.
\end{align}
The $\boldsymbol{\mu}_{\ell,m}$ represent explicit Runge-Kutta (ERK) methods of order $2m$, that is $\boldsymbol{\mu}_{\ell,m} = \boldsymbol{m}(t) + \mathcal{O}(H^{2m})$. Thus, also stability behavior is that of ERK methods. 
Here, the formula $\boldsymbol{\mu}_{\ell,m-1}$ is the most accurate approximation that can be associated with a computationally available error estimate of order $2m-1$ \cite{deuflhard1985recent} 
\begin{align}\label{errest}
 \varepsilon_{\ell,m-1} := \|\boldsymbol{\mu}_{\ell,m-1} - \boldsymbol{\mu}_{\ell,m}\| = \mathcal{O}(H^{2m-1}).
\end{align}
In practice, however, the formula $\boldsymbol{\mu}_{\ell,\ell}$ is taken as the numerical approximation of order $2\ell$ for a prescribed (or determined) level $\ell$ and the error estimate \eqref{errest} (with $m=\ell$) 
is used for step size control in the notion of local extrapolation.   
Note that, due to the different expansions \eqref{h2expansion} for odd and even indices, the corresponding orders of the error and error estimate for the extrapolated formula would be reduced by one in the case of odd step numbers $n_\ell$.\\ 
One practically relevant feature of the extrapolated Gragg method is the possibility of adapting the level $\ell$ (and hence the order) in accordance with the step size during computation. 
We will briefly describe this procedure for the order and step size adapted time integration of the LLG equation in section \ref{ordercontrol}.\\    
The extrapolation via the Aitken-Neville scheme requires, each level $\ell$, the renewed evaluation of the explicit midpoint rule with increased number of steps $n_\ell$. 
The number of right hand side evaluations up to level $\ell$ (with smoothing) is $1+\sum_{\nu=1}^\ell n_\nu$, where $F(\boldsymbol{m}^{(0)})$ is only computed for $\ell = 1$. 
This complexity is exponentially increasing in the case of the Romberg sequence, which turned out to be most effective in our tests (including step size and order control) compared to different choices, like for instance 
the harmonic sequences $\{2,4,6,8,10,12,\hdots\}$. To tame this complexity, we will treat the expensive nonlocal stray field differently from the rest of the effective field components.
\section{Taming the complexity of the extrapolation}\label{taming} 
The computational effort for derivative evaluation in numerical schemes for the LLG equation is dominated by the nonlocal part of the effective field. 
Typically, the computational effort of the stray field computation amounts to $80-90\%$ \cite{strayfield_review} of the total effective field, depending on the spatial discretization scheme, 
the numerical scheme for the stray field computation and other local field components (especially concerning the Laplacian for the exchange field). This makes plain extrapolation schemes for the LLG equation inefficient. 
On the other hand, extrapolation delivers naturally local error estimates and hence the possibility to incorporate adaptive time step selection, which is necessary to make the integration scheme practical and more robust. 
In addition, the extrapolated Gragg method offers a hierarchy of accurate higher order schemes and the opportunity to adapt the order as well. 
However, the linearity of the stray field operator with respect to the magnetization offers a way out to tame the complexity. 
We perform the full explicit midpoint scheme \eqref{Gragg} for level $\ell=1$ ($n_1 = 2,\, h_1 = H/2$), which leads to second order approximations $\boldsymbol{m}^{(1)} = \boldsymbol{m}(H/2) + \mathcal{O}(h_1^2)$ 
and $ \boldsymbol{m}^{(2)} = \boldsymbol{m}(H) + \mathcal{O}(h_1^2)$ according to \eqref{h2expansion}. Let us denote the (linear) discrete stray field operator as $\mathcal{D}$, that is $\boldsymbol{h}_s(\boldsymbol{m}) := \mathcal{D} \boldsymbol{m}$. 
We address the discretized LLG equation in the form ($t_0 = 0$)
\begin{eqnarray}\label{LLG_disc}
\begin{aligned}
 \boldsymbol{m}^\prime(t) &\, = F\big(\boldsymbol{m}(t); \mathcal{D}\big),\\
\boldsymbol{m}(0) &\,= \boldsymbol{m}^{(0)},
\end{aligned}
\end{eqnarray}
where we emphasize the dependence of the right hand side on the stray field operator $\mathcal{D}$.  
We now define a piecewise linear stray field on $[0,H/2] \cup [H/2,H]$ and denote the corresponding operator with $\mathcal{D}_H$. Hence, there shall hold the interpolation condition at exact (unknown) solution values 
\begin{align}\label{interpcond}
 \mathcal{D}_H \boldsymbol{m}(\nu h_1) &\, = \mathcal{D} \boldsymbol{m}(\nu h_1), \quad \nu = 0,1,2,
\end{align}
while for intermediate times $t\in[0,H]$ the approximation is second order.
Our discretized LLG equation \eqref{ivp} takes now the approximate form 
\begin{eqnarray}\label{LLG_approx}
\begin{aligned}
\boldsymbol{m}^\prime(t) &\, = F\big(\boldsymbol{m}(t); \mathcal{D}_H\big),\\
\boldsymbol{m}(0) &\,= \boldsymbol{m}^{(0)},
\end{aligned}
\end{eqnarray}
where $F$ depends on the linearly interpolated stray field instead of the operator $\mathcal{D}$. A computationally realizable approximation of the time-linearized operator $\mathcal{D}_H$ is first obtained 
from the computation at level $1$, where we evaluate the stray field by using the operator $\mathcal{D}$ at the iterates $\boldsymbol{m}^{(0)},\boldsymbol{m}^{(1)}$ and $\boldsymbol{m}^{(2)}$. There holds  
\begin{align}
 \mathcal{D}_H \boldsymbol{m}^{(\nu)} &\, = \mathcal{D}_H \big(\boldsymbol{m}(\nu h_1) + \mathcal{O}(h_1^2)\big) = \mathcal{D}\boldsymbol{m}(\nu h_1) + \mathcal{O}(h_1^2) = \mathcal{D}\boldsymbol{m}^{(\nu)} + \mathcal{O}(h_1^2), \quad \nu = 0,1,2.
\end{align}
Note that, due to the $h_1^2$-error expansions of the iterates \eqref{h2expansion} and the linearity of the operators $\mathcal{D}$ and $\mathcal{D}_H$, the above error also involves only even powers of $h_1$, that is
\begin{align}
 \mathcal{D}_H \boldsymbol{m}^{(\nu)} &\, = \mathcal{D}\boldsymbol{m}^{(\nu)} + \sum_{j=1}^\ell (\mathcal{D}- \mathcal{D}_H)\,\boldsymbol{e}_j(\nu h_1)\,h_1^{2j} + h_1^{2j+2} (\mathcal{D}- \mathcal{D}_H)\boldsymbol{c}(\nu h_1;h_1), \quad \nu = 0,1,2,
\end{align}
where $\boldsymbol{e}_j = \boldsymbol{a}_j$ for $\nu = 0,2$ with $\boldsymbol{a}_j(0) = 0$ and $\boldsymbol{e}_j = \boldsymbol{b}_j$ for $\nu = 1$.
This means that the interpolation conditions \eqref{interpcond} hold approximately ($\nu=1,2$) for the computational realization of $\mathcal{D}_H$ with an error involving only powers of $h_1^2$. This shall make us aware of the possibility of  
exploiting efficient extrapolation for the values $\mathcal{D} \boldsymbol{m}^{(\nu)},\,\nu = 1,2$ to establish more accurate approximations to $\mathcal{D}_H\boldsymbol{m}^{(\nu)},\,\nu = 1,2$. As step sequence we choose the Romberg sequence, 
hence the amount of steps is doubled and the step size halved  
from one to the next extrapolation level, compare with Fig.~\ref{fig:romberg}. Each level we perform the Gragg method \eqref{Gragg} with smoothing, where we save stray field evaluations by using interpolated values from the current approximate version of $\mathcal{D}_H$, i.e. 
we are solving \eqref{LLG_approx}. 
Renewed evaluations of the stray field are only necessary at $t=H/2$ and $t=H$ followed by the computation of a new line in the Aitken-Neville scheme \eqref{AitkenNeville} for both, 
the current iterate $\boldsymbol{m}_{h_\ell}(H)$ and the stray field values  $\mathcal{D} \boldsymbol{m}^{(n_{\ell}/2)}$ and $\mathcal{D} \boldsymbol{m}^{(n_{\ell})}$. Note  
that the midpoint in level $1$ has odd parity, while for all subsequent levels it is even. We therefore use for $\ell=1$ a centered average of the stray field at the midpoint. This is 
\begin{align}
 \boldsymbol{m}_1^{(n_{1}/2)} &\,:= \frac{1}{2}\big(\boldsymbol{m}^{(n_{1}/2+1)} + \boldsymbol{m}^{(n_{1}/2-1)}\big) = \boldsymbol{m}^{(n_{1}/2)} + \mathcal{O}(h_1^2),
\end{align}
where the error term involves only even powers of $h_1$.
Note that no further evaluations of $\mathcal{D}$ are needed here, since 
\begin{align}
 \mathcal{D}\boldsymbol{m}_1^{(n_{1}/2)} &\,= \frac{1}{2}\big(\mathcal{D}\boldsymbol{m}^{(n_{1}/2+1)} + \mathcal{D}\boldsymbol{m}^{(n_{1}/2-1)}\big),
\end{align}
where the values on the right hand side are already available.
\section{Step size and order control}\label{ordercontrol}
According to \cite{hairer,deuflhard1985recent} we take for level $\ell$ the order $2\ell-1$ error estimate \eqref{errest} of $\boldsymbol{\mu}_{\ell,\ell-1}$ for the numerical approximation $\boldsymbol{\mu}_{\ell,\ell}$ in the 
notion of local extrapolation. We remark, that also the error estimates of the extrapolation of the stray field values are available and can be incorporated in several different ways.  
One possibility is to simply establish a weighted sum of relative error estimates.    
As usual, we require the dominant term $err_\ell \approx C\,H^{2\ell-1}$ in the error estimate for a given basic step size $H$ to reach a tolerance $tol \approx C \widetilde{H}^{2\ell-1}$ obtained 
from an adapted step size $\widetilde{H}$. This, together with incorporated safety factors, yields the empirically optimal choice \cite{hairer_old} for an adapted step size $H_\ell$ at level $\ell$
\begin{align}\label{adaptH}
 H_\ell = 0.94\cdot H\cdot\Big(0.65\, \frac{tol}{err_\ell}\Big)^{1/(2\ell -1)}.  
\end{align}
Equation \eqref{adaptH} is used for determining a next step size within a convergence monitor for the three subsequent levels $\ell,\ell+1$ and $\ell+2$, which determine whether the current approximation is accepted or rejected and the order increased or decreased \cite{hairer_old}.   
The tool for measuring the necessity and efficiency for order and step size adaption is the reduction of work per time step size $W_\ell/H_\ell$, where the work $W_\ell$ measures the effort for computing the numerical approximation $\boldsymbol{\mu}_{\ell,\ell}$. 
An adapted choice for the step size and order shall reduce the work per time step size. 
We define $W_\ell$ as the weighted sum of stray field evaluations and effective field evaluations up to level $\ell$. The latter one uses already computed stray field evaluations and, 
hence, can be understood as the amount of evaluations of all other field components except the stray field. As a weighting factor we take $f_{sf} = 0.8-0.9$, which shall be a rough estimate of 
the portion of the costs for the stray field compared to the total field. Hence, the work is defined as
\begin{align}
 W_\ell = f_{sf}\, (2\ell+1) + (1-f_{sf})\, (1+\sum_{\nu=1}^\ell n_\nu) = f_{sf}\, (2\ell+1) + (1-f_{sf})\, (2^{\ell+1}-1),
\end{align}
where $n_\nu = 2^\nu$ for the Romberg sequence. Note that the stray field part of the work increases linearly with the level, while the other part increases exponentially.   
\section{Numerics}
\begin{figure}[hbtp]
\centering 
\includegraphics[scale=0.38]{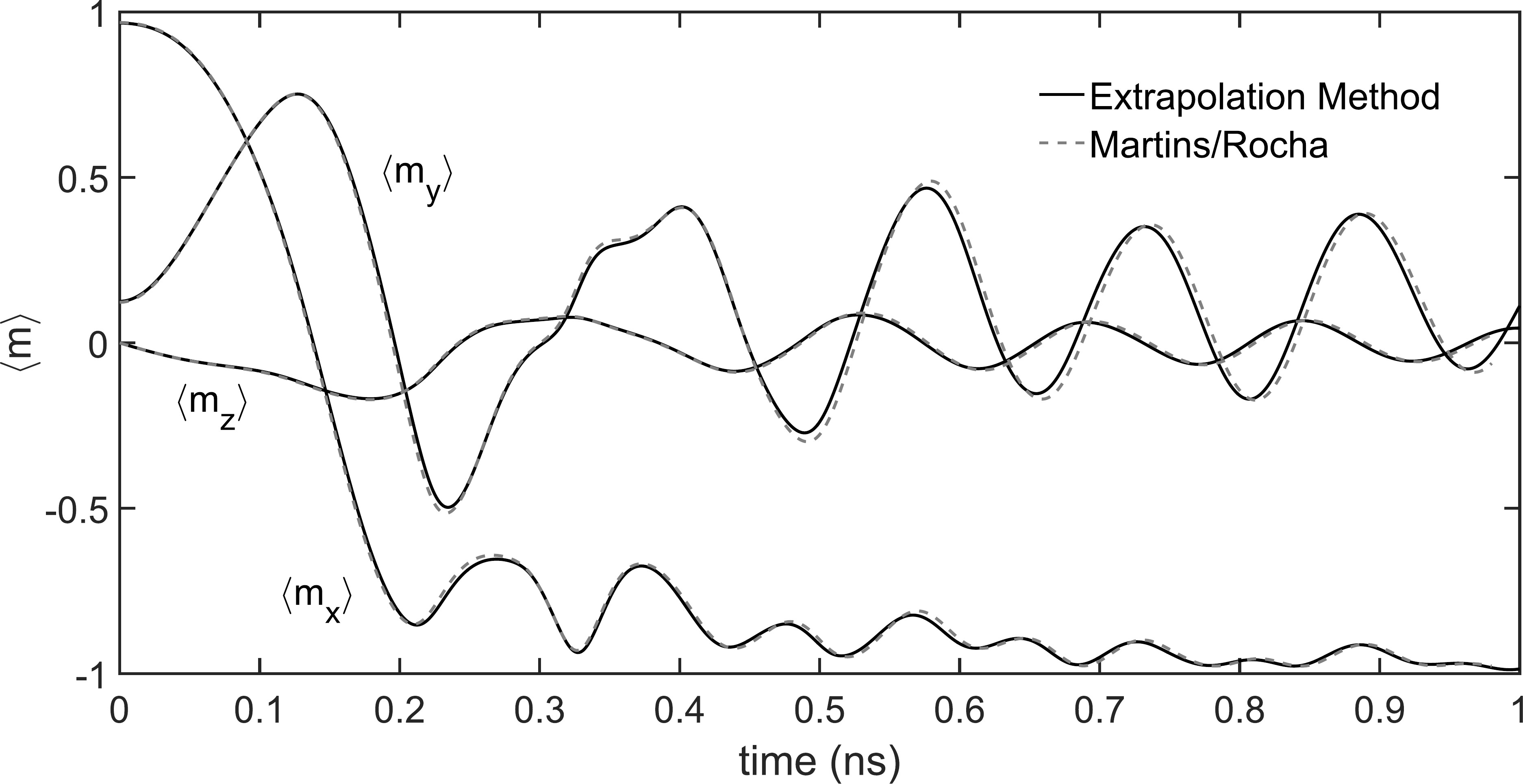}    
\caption{Time evolution of averaged magnetization components for $\mu$MAG Standard problem $\#4$ (first external field) with $1$nm discretization and $tol = 1.0 \times 10^{-12}$. Comparison with published results of Martins/Rocha for their $1$nm computation \cite{mumag4}.}\label{fig:SP4_vgl1}
\end{figure}

\begin{figure}[hbtp]
\centering 
\includegraphics[scale=0.33]{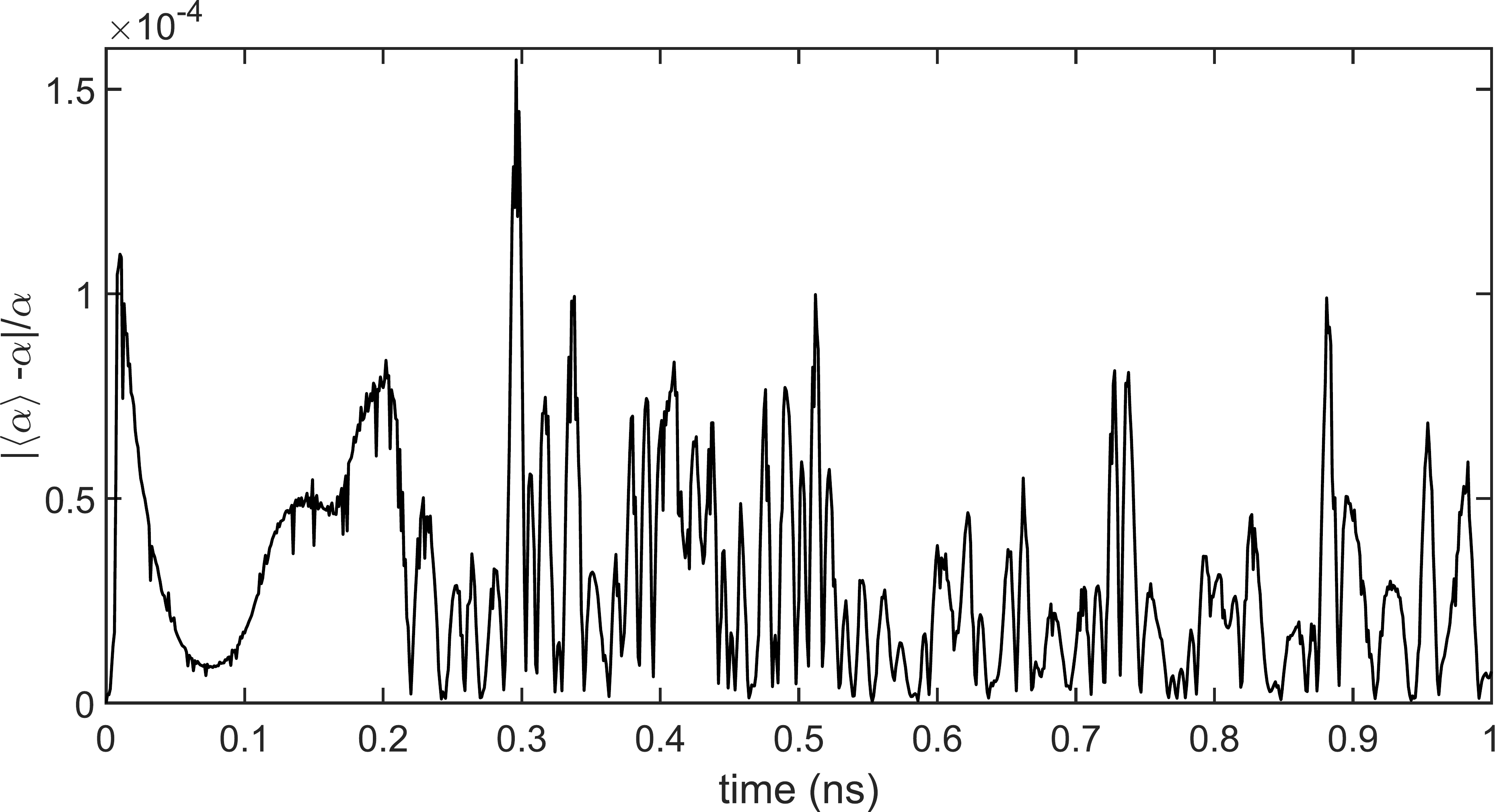}    
\caption{Relative errors $|\langle\alpha\rangle - \alpha|/\alpha$ of the approximated numerical damping parameter during time propagation in Fig.~\ref{fig:SP4_vgl1}.}\label{fig:SP4_vgl1_alpha}
\end{figure}

\begin{figure}[hbtp]
\centering 
\includegraphics[scale=0.38]{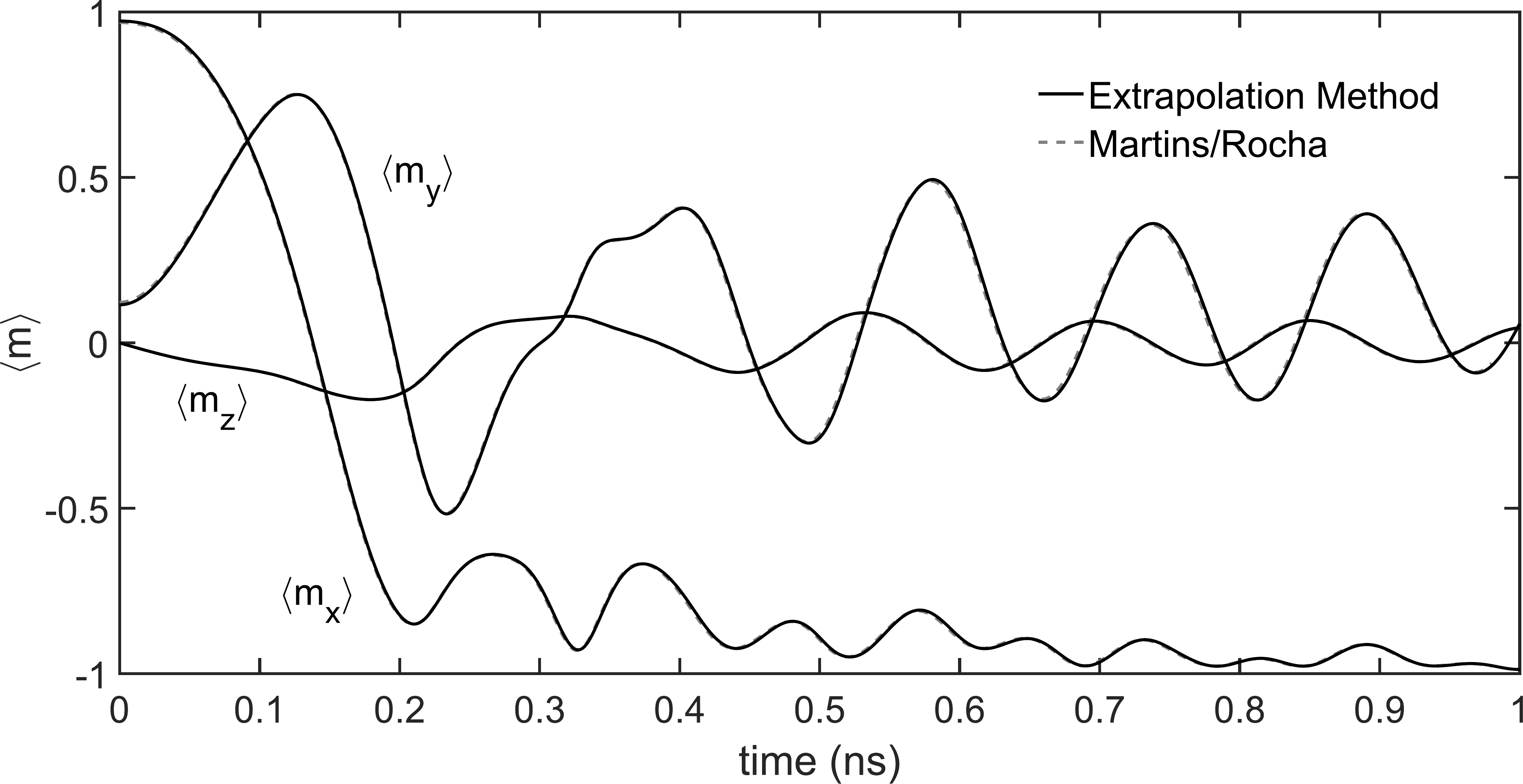}    
\caption{Time evolution of averaged magnetization components for $\mu$MAG Standard problem $\#4$ (first external field) with discretization consisting of $2 \times 2 \times 1$nm prisms and $tol = 1.0 \times 10^{-12}$. Comparison with published results of Martins/Rocha for their $1$nm computation \cite{mumag4}.}\label{fig:SP4_vgl2}
\end{figure}

We look at the NIST $\mu$MAG Standard problem $\#4$ \cite{mumag4}. The geometry is a magnetic plate of size $500 \times 125 \times 3$ nm with material parameters of permalloy: 
$A = 1.3 \times 10^{-11}$ J/m, $M_s = 8.0 \times 10^5$ A/m, $\alpha = 0.02$. The initial state is an equilibrium s-state, obtained after applying and slowly reducing a saturating field along the diagonal direction $[1,1,1]$ to zero. 
Then an external field of magnitude $25$mA is applied with an angle of $170^\circ$ c.c.w. from the positive $x$ axis. We use different spatial discretizations, where the finest is built of $1$nm cubes. Errors are computed in the 
relative Euclidean norm, where we weighted error estimates from the stray field extrapolation marginally with two percent. However, investigation of the decrease of the error estimates of the stray field values at the mid- and endpoint, 
$H/2$ and $H$ respectively, showed analogue decay rate and magnitudes as for the error estimates of the magnetization iterates at the endpoint. 
Simulations are performed by subdividing the time interval into $1$ps subintervals and data were captured at every $1$ps simulated time. Some measures like the time step sizes, the extrapolation level or the numerical 
damping parameter were recorded within the $1$ps subintervals and archived as averaged values. Fig.~\ref{fig:SP4_vgl1} shows the time evolution of the averaged magnetization components for $1$ns simulated time and 
the $1$nm cube discretization. Computations in Fig.~\ref{fig:SP4_vgl1} were performed with $f_{sf} = 0.85$ and a tolerance of $tol = 1.0 \times 10^{-12}$. 
In Tab.~\ref{tab:SP4_vgl1} we give statistics of these computations for tolerances of $1.0\times 10^{-10}$ and $1.0\times10^{-12}$ including the spent work w.r.t. $f_{sf} = 0.8,0.85$ and $0.9$, number of function evaluations, 
average extrapolation level and step size, number of rejected steps and absolute maximum error of unit norm constraint (LLG preserves the magnitude of the moments). No renormalization is performed.
\begin{table}[h!]
\tabcolsep 1.2pt \caption{Statistics of $1$ns simulation of $\mu$MAG Standard problem $\#4$ on a grid consisting of $1$nm cubes for tolerances of $1.0\times 10^{-10}$ and $1.0\times10^{-12}$ giving the spent work $W(f_{sf})$ w.r.t. $f_{sf} = 0.8,0.85$ and $0.9$, 
number of stray field evaluations $\#\texttt{fevs}$, average order $\langle\texttt{level}\rangle$ and average step size $\langle\Delta t\rangle$ (in fs), number of rejected steps $\#\texttt{rej}$ and 
absolute maximum error of unit norm constraint $\texttt{errnm}$. Our method is abbreviated with 
\texttt{ExMP} (Extrapolated MidPoint) and the Runge-Kutta method with \texttt{DP87}.} \label{tab:SP4_vgl1}
\begin{center}\vspace{-1em}
\def\temptablewidth{16cm}
{\rule{\temptablewidth}{1.3pt}}
\begin{tabularx}{16cm}{{@{}X *9{>{\centering\arraybackslash}X}@{}}}
Method         & $\texttt{tol}$ & $W(0.90)$ &  $W(0.85)$ & $W(0.80)$ & $\#\texttt{fevs}$ & $\langle\texttt{level}\rangle$ & $\langle\Delta t\rangle$  & $\#\texttt{rej}$ & $\texttt{errnm}$ \\ \hline\hline
\texttt{ExMP}  & 1E-12          & 131338    &  162983    & 194629    &   68047           &          6.106                 &           195.8           &  0/5167          & 3.6E-13          \\ \hline  
\texttt{ExMP}  & 1E-10          & 130601    &  162233    & 193865    &   67337           &          6.118                 &           197.3           &  0/5085          & 1.6E-10          \\ \hline
\texttt{DP87}  & 1E-10          &      -    &       -    &      -    &   197080          &             -                  &           67.5            &  99/14830        & 1.1E-13          \\ \hline
\texttt{DP87}  & 1E-08          &      -    &       -    &      -    &   197296          &             -                  &           67.7            &  114/14797       & 9.3E-10          \\ \hline
\end{tabularx}
{\rule{\temptablewidth}{1.3pt}}
\end{center}
\end{table}
We also compared our method with the \texttt{Dopri8} (Dormand-Prince) method from \cite{hairer_old}, which is an $8$th order embedded Runge-Kutta method using a $7$th order estimate for the error, 
results are also included in Tab.~\ref{tab:SP4_vgl1}. 
Averaged magnetization components coincide with our method with absolute error in the range of $1.0\times 10^{-6} - 1.0\times 10^{-4}$. At time $t=0.138$ns, approximately the moment where $\langle m_x \rangle$ crosses zero for the first time, the 
discrete magnetization configurations were captured for the Runge-Kutta and our method, giving a calculated maximum 
absolute deviation on the entire $1$nm grid of about $1.0 \times 10^{-6}$. 
\begin{table}[h!]
\tabcolsep 1.2pt \caption{Statistics of $1$ns simulation of $\mu$MAG Standard problem $\#4$ on a $250\times64\times3$ grid for tolerances of $1.0\times 10^{-10}$ and $1.0\times10^{-12}$ giving the spent work $W(f_{sf})$ w.r.t. $f_{sf} = 0.8,0.85$ and $0.9$, 
number of stray field evaluations $\#\texttt{fevs}$, average order $\langle\texttt{level}\rangle$ and average step size $\langle\Delta t\rangle$ (in fs), number of rejected steps $\#\texttt{rej}$ and 
absolute maximum error of unit norm constraint $\texttt{errnm}$. Our method is abbreviated with 
\texttt{ExMP} (Extrapolated MidPoint) and the Runge-Kutta method with \texttt{DP87}.} \label{tab:SP4_vgl2}
\begin{center}\vspace{-1em}
\def\temptablewidth{16cm}
{\rule{\temptablewidth}{1.3pt}}
\begin{tabularx}{16cm}{{@{}X *9{>{\centering\arraybackslash}X}@{}}}
Method         & $\texttt{tol}$ & $W(0.90)$ &  $W(0.85)$ & $W(0.80)$ & $\#\texttt{fevs}$ & $\langle\texttt{level}\rangle$ & $\langle\Delta t\rangle$  & $\#\texttt{rej}$ & $\texttt{errnm}$ \\ \hline\hline
\texttt{ExMP}  & 1E-12          & 81142     &   99281    & 117420    &   44864           &          5.946                 &           301.9           &  0/3550          & 1.1E-13          \\ \hline  
\texttt{ExMP}  & 1E-10          & 71990     &    89347   &   106705  &   37275           &          6.107                 &           363.5           &  0/2829          & 1.0E-10          \\ \hline
\texttt{DP87}  & 1E-10          &      -    &       -    &      -    &   108888          &             -                  &           124.9           &  370/8006        & 6.6E-14          \\ \hline
\texttt{DP87}  & 1E-08          &      -    &       -    &      -    &   107991          &             -                  &           125.5           &  336/7971        & 8.2E-11          \\ \hline
\end{tabularx}
{\rule{\temptablewidth}{1.3pt}}
\end{center}
\end{table}
We consider one more error measure: the relative error of the numerical damping parameter $|\langle\alpha\rangle - \alpha|/\alpha$ at a time $t_\nu$ according to \cite{miltat2007numerical} 
\begin{align}\label{alpha_est}
 \langle\alpha\rangle = -h \frac{\sum_j \Delta \varepsilon_j^{(\nu)} }{\sum_j |\Delta \boldsymbol{m}_j^{(\nu)}|^2},
\end{align}
where $j$ is the node index and we approximate $\Delta \varepsilon_j^{(\nu)} = \varepsilon^{(\nu)}_j-\varepsilon^{(\nu-1)}_j \approx  -\boldsymbol{h}^{(\nu-1/2)}_j\cdot\Delta \boldsymbol{m}_j^{(\nu)} = -\boldsymbol{h}^{(\nu-1/2)}_j\cdot (\boldsymbol{m}^{(\nu)}_j-\boldsymbol{m}^{(\nu-1)}_{j})$ and 
use for the field at the midpoint $\boldsymbol{h}^{(\nu-1/2)}_j \approx (\boldsymbol{h}^{(\nu)}_j + \boldsymbol{h}^{(\nu-1)}_j)/2$. Note, however, that this is itself a first order approximation to the analytical expression
\begin{align}
 \alpha = - \frac{\int_\Omega \text{d} \varepsilon/\text{d} \tau}{\int_\Omega (\text{d}\boldsymbol{m}/\text{d}\tau)^2},
\end{align}
where $\varepsilon$ is the energy density. The errors are plotted in Fig.~\ref{fig:SP4_vgl1_alpha} associated with the computations of Fig.~\ref{fig:SP4_vgl1}, where the average step size was about $195$fs. 
The computations in Tab.~\ref{tab:SP4_vgl1} are repeated on a coarser mesh consisting of about $2 \times 2 \times 1$nm prisms (mesh size $250\times64\times3$), see Fig.~\ref{fig:SP4_vgl2} and Tab.~\ref{tab:SP4_vgl2}.
\begin{figure}[hbtp]
\centering 
\includegraphics[scale=0.43]{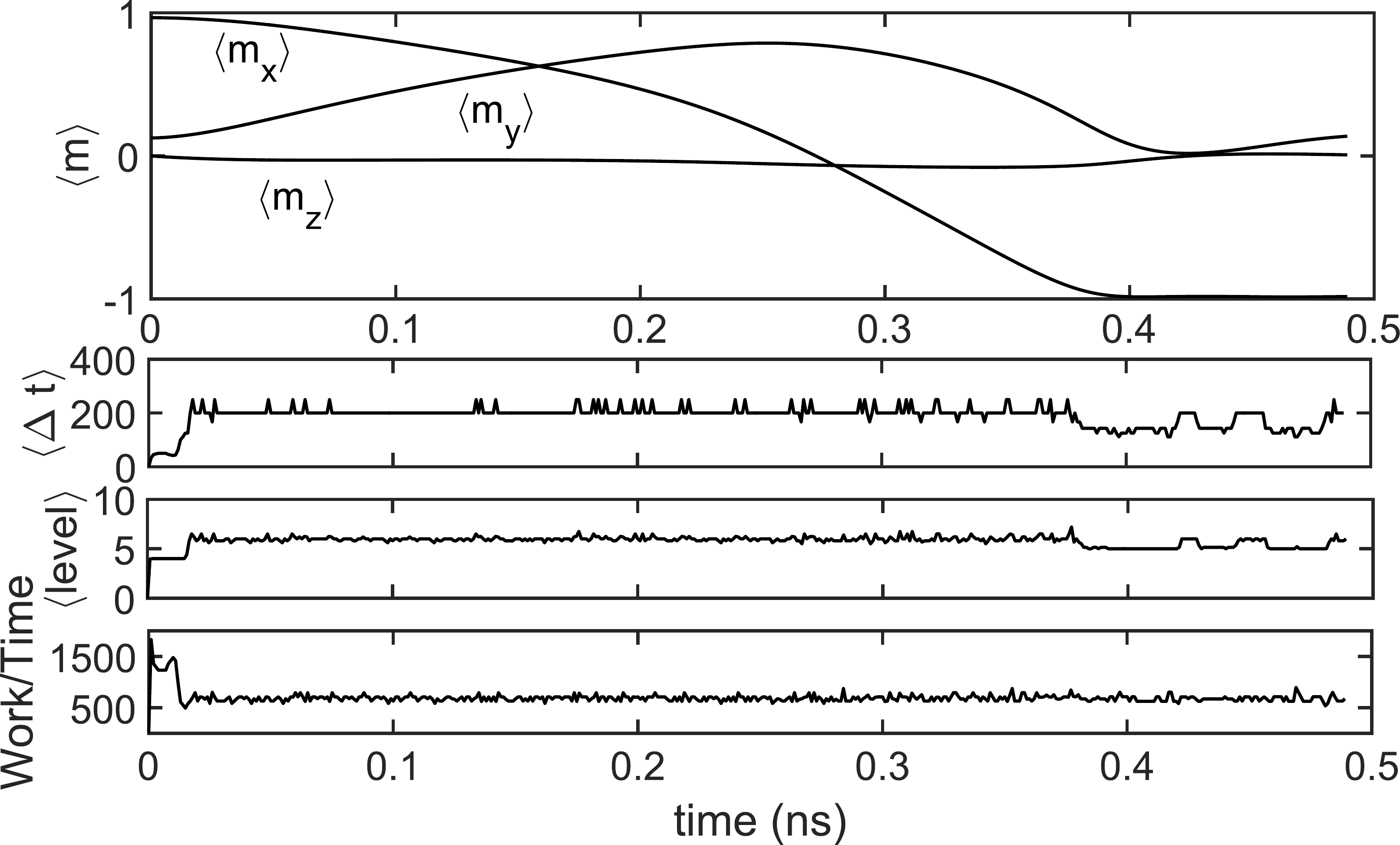}    
\caption{Same parameters as for Fig.~\ref{fig:SP4_vgl1} but $\alpha = 0.2$. Propagation of averaged magnetization $\langle m\rangle$ with $1$nm discretization, the averaged time steps $\langle\Delta t\rangle$ in fs, 
the average extrapolation levels $\langle\texttt{level}\rangle$ and work per (reduced) time step.}\label{fig:SP4alpha02}
\end{figure}
\begin{figure}[hbtp]
\centering 
\includegraphics[scale=0.36]{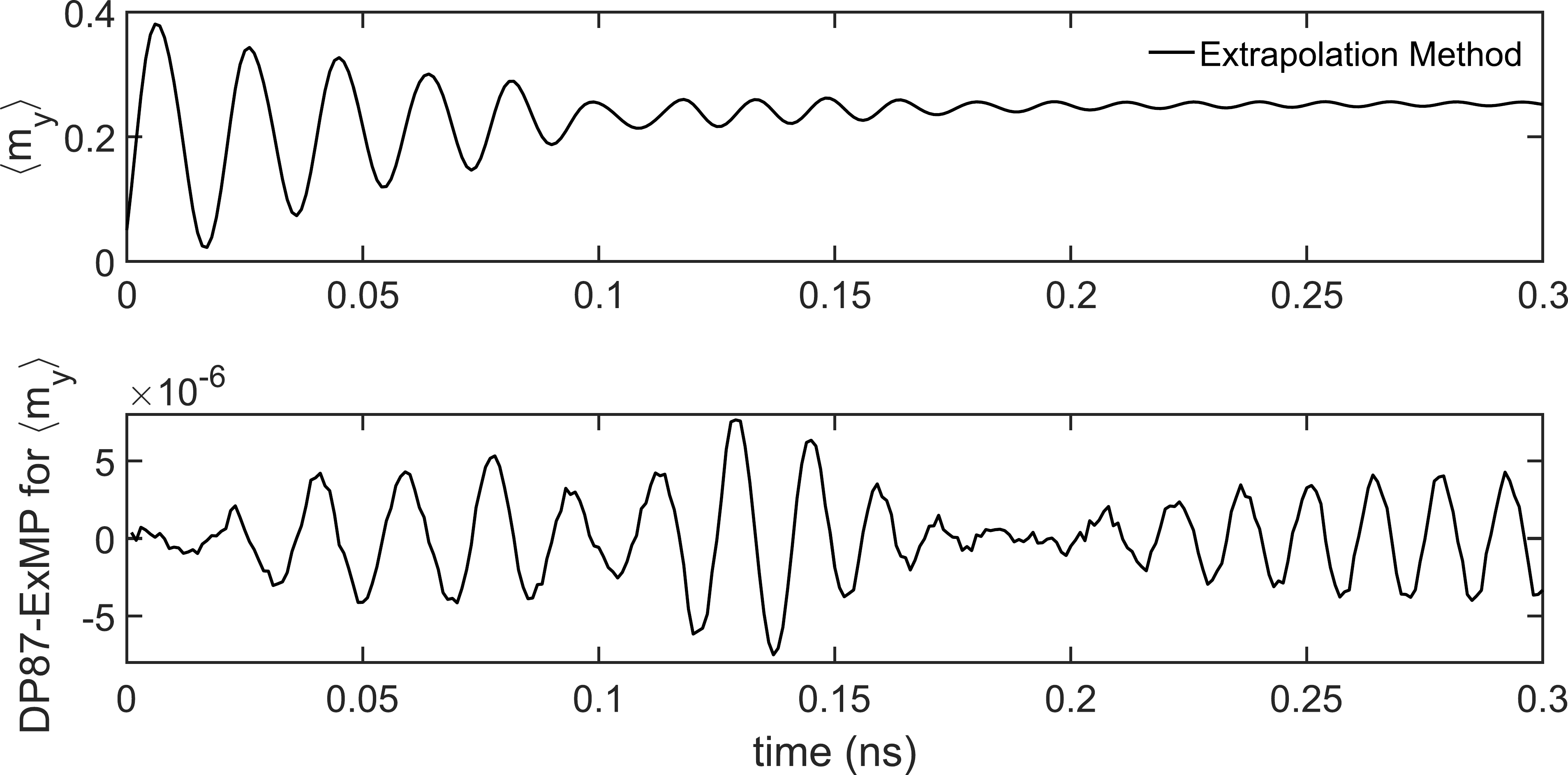}    
\caption{$K_1 = 1.0 \times 10^{6}$ $J/m^3$, easy axis $\boldsymbol{a} = [1,1,1]/\sqrt{3}$ and $1$nm discretization. Propagation of the averaged $y$-component and deviation from the corresponding values obtained from the Runge-Kutta method.}\label{fig:SP4Q124_vgl1}
\end{figure}
\begin{figure}[hbtp]
\centering 
\includegraphics[scale=0.36]{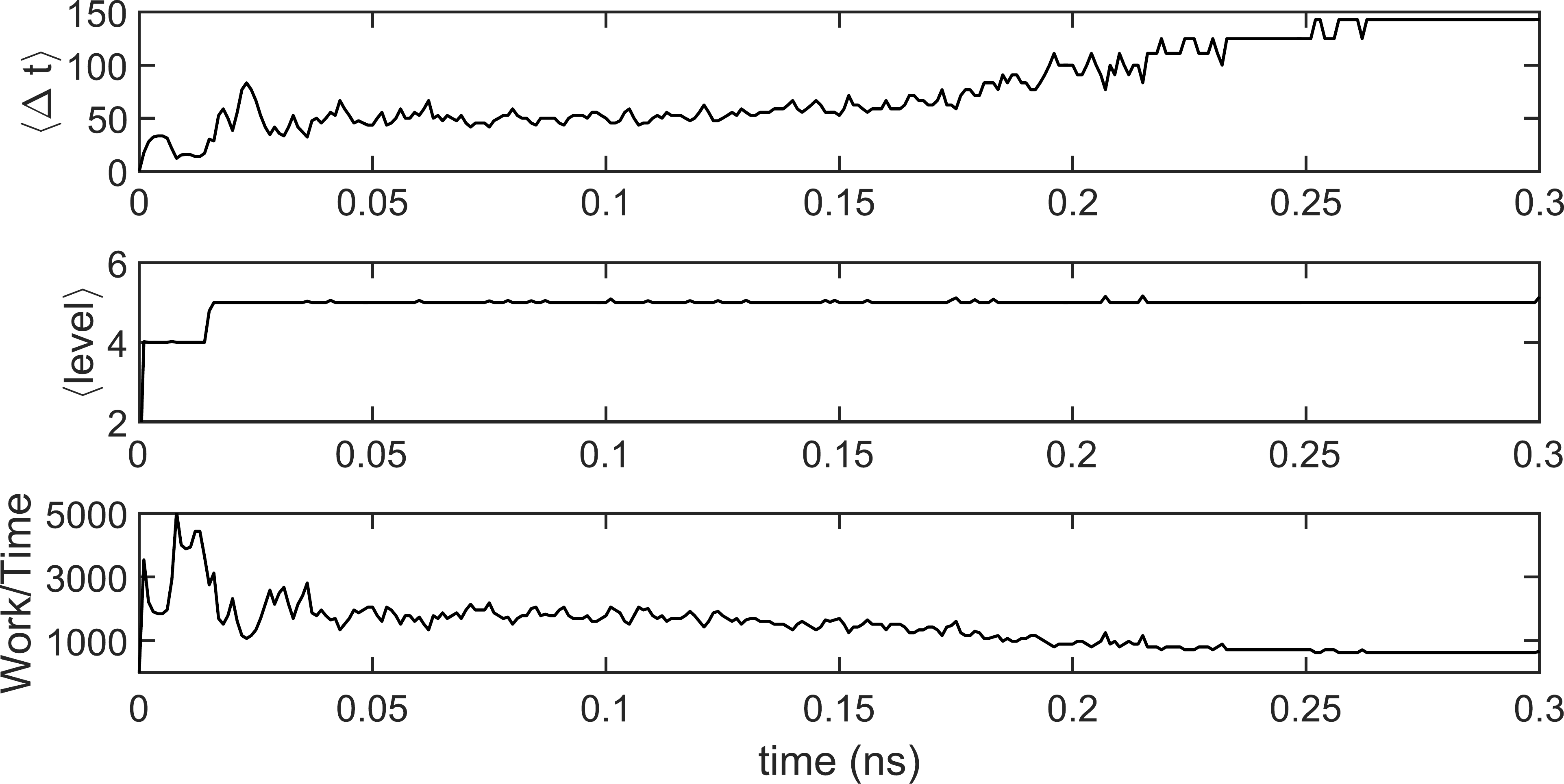}    
\caption{$K_1 = 1.0 \times 10^{6}$ J/m$^3$, easy axis $\boldsymbol{a} = [1,1,1]/\sqrt{3}$ and $1$nm discretization. Averaged time steps $\langle\Delta t\rangle$ in fs, 
the average extrapolation levels $\langle\texttt{level}\rangle$ and work per (reduced) time step.}\label{fig:SP4Q124_vgl2}
\end{figure}%
As a second numerical test we change parameters of the original setting of Standard problem $\#4$. 
Results in Fig.\ref{fig:SP4alpha02} were obtained by changing the damping parameter to $\alpha = 0.2$ and shows propagation of averaged magnetization with $1$nm discretization, the averaged time steps (within the $1$ps subintervals), 
the average extrapolation levels and work per (reduced) time step. One can recognize the interplay between order and step size adaption, while the work per time step remains roughly unchanged.\\  
Now we change the anisotropy constant to $K_1 = 1.0 \times 10^{6}$ J/m$^3$ with the easy axis $\boldsymbol{a} = [1,1,1]/\sqrt{3}$, while maintaining all other original parameters. Fig.~\ref{fig:SP4Q124_vgl1} shows computation results with $1$nm discretization for 
the propagation of the averaged $y$-component (the others oscillate similarly) and the deviation from the corresponding values obtained from the Runge-Kutta method. In Fig.~\ref{fig:SP4Q124_vgl2} we give averaged time steps (within the $1$ps subintervals), 
the average extrapolation levels and work per (reduced) time step, associated with Fig.~\ref{fig:SP4Q124_vgl1}.\\
From the test examples one can recognize an advantage of the extrapolation method in terms of required work. This aspect gets more significant when the true $f_{sf}$ gets larger, that is, stray field computation increasingly dominates 
other computational costs. In all tests the average step sizes are clearly larger and stray field computations are fewer. Moreover, it is noticeable that due to 
the simultaneous order and step size control there are actually no rejected (wasted) steps.  
\newpage
\section{Conclusions}
We developed a step size and order adaptive solver for the Landau-Lifschitz-Gilbert equation. The method uses extrapolation of the symmetric explicit midpoint scheme, which possess an asymptotic error expansion in even powers of the step size 
parameter. The necessary number of expensive stray field evaluations is reduced to linear dependence on the order of the method. This is achieved by a piecewise time-linear stray field approximation. 
We show how to efficiently extrapolate this approximation by utilizing the $h^2$-expansion and the linearity of the stray field operator. 
Numerical experiments indicate that the proposed scheme gets more and more efficient, compared to conventional methods as higher order Runge-Kutta, 
when the stray field computation takes a larger portion of the costs for the effective field evaluation. This is more likely the case in field-based stray field approaches.

\section*{Acknowledgments}
Financial support by the Austrian Science Foundation (FWF) under grant No F41 (SFB 'VICOM'), grant No F65 (SFB 'Complexity in PDEs') and grant No W1245 (DK 'Nonlinear PDEs') and the Wiener Wissenschafts- und TechnologieFonds (WWTF) project  No MA16-066 ('SEQUEX'). 
The computational results have been achieved using the Vienna Scientific Cluster (VSC).

\bibliographystyle{unsrtnat}
\bibliography{refs}

\begin{thebibliography}{29}
\providecommand{\natexlab}[1]{#1}
\providecommand{\url}[1]{\texttt{#1}}
\expandafter\ifx\csname urlstyle\endcsname\relax
  \providecommand{\doi}[1]{doi: #1}\else
  \providecommand{\doi}{doi: \begingroup \urlstyle{rm}\Url}\fi

\bibitem[Brown(1963)]{brown1963micromagnetics}
W.~F. Brown.
\newblock \emph{Micromagnetics}.
\newblock Number~18. Interscience Publishers, 1963.

\bibitem[Suess et~al.(2015)Suess, Vogler, Abert, Bruckner, Windl, Breth, and
  Fidler]{suess2015fundamental}
D.~Suess, C.~Vogler, C.~Abert, F.~Bruckner, R.~Windl, L.~Breth, and J.~Fidler.
\newblock Fundamental limits in heat-assisted magnetic recording and methods to
  overcome it with exchange spring structures.
\newblock \emph{Journal of Applied Physics}, 117\penalty0 (16):\penalty0
  163913, 2015.

\bibitem[Kovacs et~al.(2016)Kovacs, Oezelt, Schabes, and
  Schrefl]{kovacs2016numerical}
A.~Kovacs, H.~Oezelt, M.E. Schabes, and T.~Schrefl.
\newblock Numerical optimization of writer and media for bit patterned magnetic
  recording.
\newblock \emph{Journal of Applied Physics}, 120\penalty0 (1):\penalty0 013902,
  2016.

\bibitem[Makarov et~al.(2012)Makarov, Sverdlov, Osintsev, and
  Selberherr]{makarov2012fast}
A.~Makarov, V.~Sverdlov, D.~Osintsev, and S.~Selberherr.
\newblock Fast switching in magnetic tunnel junctions with two pinned layers:
  Micromagnetic modeling.
\newblock \emph{IEEE Transactions on Magnetics}, 48\penalty0 (4):\penalty0
  1289--1292, 2012.

\bibitem[Sepehri-Amin et~al.(2013)Sepehri-Amin, Ohkubo, Nagashima, Yano, Shoji,
  Kato, Schrefl, and Hono]{sepehri2013high}
H.~Sepehri-Amin, T.~Ohkubo, S.~Nagashima, M.~Yano, T.~Shoji, A.~Kato,
  T.~Schrefl, and K.~Hono.
\newblock High-coercivity ultrafine-grained anisotropic nd--fe--b magnets
  processed by hot deformation and the nd--cu grain boundary diffusion process.
\newblock \emph{Acta Materialia}, 61\penalty0 (17):\penalty0 6622--6634, 2013.

\bibitem[Bance et~al.(2014)Bance, Oezelt, Schrefl, Winklhofer, Hrkac, Zimanyi,
  Gutfleisch, Evans, Chantrell, Shoji, Yano, Sakuma, Kato, and
  Manabe]{bance2014high}
S.~Bance, H.~Oezelt, T.~Schrefl, M.~Winklhofer, G.~Hrkac, G.~Zimanyi,
  O.~Gutfleisch, R.F.L. Evans, R.W. Chantrell, T.~Shoji, M.~Yano, N.~Sakuma,
  A.~Kato, and A.~Manabe.
\newblock High energy product in battenberg structured magnets.
\newblock \emph{Applied Physics Letters}, 105\penalty0 (19):\penalty0 192401,
  2014.

\bibitem[Kronmüller(2007)]{kronmueller}
H.~Kronmüller.
\newblock \emph{General Micromagnetic Theory}.
\newblock John Wiley \& Sons, Ltd, 2007.
\newblock ISBN 9780470022184.
\newblock \doi{10.1002/9780470022184.hmm201}.
\newblock URL \url{http://dx.doi.org/10.1002/9780470022184.hmm201}.

\bibitem[d’Aquino et~al.(2005)d’Aquino, Serpico, and
  Miano]{d2005geometrical}
M.~d’Aquino, C.~Serpico, and G.~Miano.
\newblock Geometrical integration of {L}andau--{L}ifshitz--{G}ilbert equation
  based on the mid-point rule.
\newblock \emph{Journal of Computational Physics}, 209\penalty0 (2):\penalty0
  730--753, 2005.

\bibitem[Suess et~al.(2002)Suess, Tsiantos, Schrefl, Fidler, Scholz, Forster,
  Dittrich, and Miles]{suess2002time}
D.~Suess, V.~Tsiantos, T.~Schrefl, J.~Fidler, W.~Scholz, H.~Forster,
  R.~Dittrich, and J.J. Miles.
\newblock Time resolved micromagnetics using a preconditioned time integration
  method.
\newblock \emph{Journal of Magnetism and Magnetic Materials}, 248\penalty0
  (2):\penalty0 298--311, 2002.
\newblock URL \url{http://dx.doi.org/10.1016/S0304-8853(02)00341-4}.

\bibitem[Donahue and Porter(1999)]{donahue1999oommf}
M.~J. Donahue and D.~G. Porter.
\newblock Oommf user’s guide, version 1.0, interagency report nistir 6376.
\newblock \emph{National Institute of Standards and Technology}, 1999.

\bibitem[Alouges and Jaisson(2006)]{alouges2006convergence}
F.~Alouges and P.~Jaisson.
\newblock Convergence of a finite element discretization for the
  landau--lifshitz equations in micromagnetism.
\newblock \emph{Mathematical Models and Methods in Applied Sciences},
  16\penalty0 (02):\penalty0 299--316, 2006.

\bibitem[Bartels and Prohl(2006)]{bartels2006convergence}
S.~Bartels and A.~Prohl.
\newblock Convergence of an implicit finite element method for the
  landau--lifshitz--gilbert equation.
\newblock \emph{SIAM journal on numerical analysis}, 44\penalty0 (4):\penalty0
  1405--1419, 2006.

\bibitem[Kritsikis et~al.(2014)Kritsikis, Vaysset, Buda-Prejbeanu, Alouges, and
  Toussaint]{kritsikis2014beyond}
E.~Kritsikis, A.~Vaysset, L.~D. Buda-Prejbeanu, F.~Alouges, and J.-C.
  Toussaint.
\newblock Beyond first-order finite element schemes in micromagnetics.
\newblock \emph{Journal of Computational Physics}, 256:\penalty0 357--366,
  2014.

\bibitem[Abert et~al.(2013)Abert, Exl, Selke, Drews, and
  Schrefl]{strayfield_review}
C.~Abert, L.~Exl, G.~Selke, A.~Drews, and T.~Schrefl.
\newblock Numerical methods for the stray-field calculation: A comparison of
  recently developed algorithms.
\newblock \emph{Journal of Magnetism and Magnetic Materials}, 326:\penalty0
  176--185, 2013.

\bibitem[Exl et~al.(2014{\natexlab{a}})Exl, Bance, Reichel, Schrefl, Stimming,
  and Mauser]{exl2014labonte}
L.~Exl, S.~Bance, F.~Reichel, T.~Schrefl, H.-P. Stimming, and N.~J. Mauser.
\newblock La{B}onte's method revisited: An effective steepest descent method
  for micromagnetic energy minimization.
\newblock \emph{Journal of Applied Physics}, 115\penalty0 (17):\penalty0
  17D118, 2014{\natexlab{a}}.

\bibitem[Fischbacher et~al.(2017)Fischbacher, Kovacs, Oezelt, Schrefl, Exl,
  Fidler, Suess, Sakuma, Yano, Kato, Shoji, and
  Manabe]{fischbacher2017conjugate}
J.~Fischbacher, A.~Kovacs, H.~Oezelt, T.~Schrefl, L.~Exl, J.~Fidler, D.~Suess,
  N.~Sakuma, M.~Yano, A.~Kato, T.~Shoji, and A.~Manabe.
\newblock Conjugate gradient methods in micromagnetics.
\newblock \emph{arXiv preprint arXiv:1701.05810}, 2017.

\bibitem[Garcia-Cervera(2007)]{garcia2007numerical}
C.~J. Garcia-Cervera.
\newblock Numerical micromagnetics: A review.
\newblock \emph{Boc. Soc. Esp. Mat. Apl.}, 39\penalty0 (103--135), 2007.

\bibitem[Hairer et~al.(1987)Hairer, N{\o}rsett, and Wanner]{hairer_old}
E.~Hairer, S.~P. N{\o}rsett, and G.~Wanner.
\newblock \emph{Solving Ordinary Differential Equations I (Nonstiff problems)}.
\newblock Springer Series in Computational Mathematics, 1987.
\newblock ISBN 978-3-662-12607-3.
\newblock \doi{10.1007/978-3-662-12607-3}.
\newblock URL \url{http://dx.doi.org/10.1007/978-3-662-12607-3}.

\bibitem[Tsiantos et~al.(2001)Tsiantos, Suess, Schrefl, and
  Fidler]{tsiantos2001stiffness}
V.~D. Tsiantos, D.~Suess, T.~Schrefl, and J.~Fidler.
\newblock Stiffness analysis for the micromagnetic standard problem no. 4.
\newblock \emph{Journal of Applied Physics}, 89\penalty0 (11):\penalty0
  7600--7602, 2001.

\bibitem[Gragg(1965)]{gragg}
W.~B. Gragg.
\newblock On extrapolation algorithms for ordinary initial value problems.
\newblock \emph{Journal of the Society for Industrial and Applied Mathematics,
  Series B: Numerical Analysis}, 2\penalty0 (3):\penalty0 384--403, 1965.
\newblock \doi{10.1137/0702030}.
\newblock URL \url{http://dx.doi.org/10.1137/0702030}.

\bibitem[Stetter(1970)]{stetter1970symmetric}
H.~J. Stetter.
\newblock Symmetric two-step algorithms for ordinary differential equations.
\newblock \emph{Computing}, 5\penalty0 (3):\penalty0 267--280, 1970.

\bibitem[Deuflhard(1985)]{deuflhard1985recent}
P.~Deuflhard.
\newblock Recent progress in extrapolation methods for ordinary differential
  equations.
\newblock \emph{SIAM review}, 27\penalty0 (4):\penalty0 505--535, 1985.

\bibitem[Bulirsch and Stoer(1966)]{bulirsch1966numerical}
R.~Bulirsch and J.~Stoer.
\newblock Numerical treatment of ordinary differential equations by
  extrapolation methods.
\newblock \emph{Numerische Mathematik}, 8\penalty0 (1):\penalty0 1--13, 1966.

\bibitem[Hairer et~al.(1993)Hairer, N{\o}rsett, and Wanner]{hairer}
E.~Hairer, S.~P. N{\o}rsett, and G.~Wanner.
\newblock \emph{Solving Ordinary Differential Equations I (Nonstiff problems)}.
\newblock Springer Series in Computational Mathematics, 1993.
\newblock ISBN 978-3-540-56670-0.
\newblock \doi{10.1007/978-3-540-78862-1}.
\newblock URL \url{http://dx.doi.org/10.1007/978-3-540-78862-1}.

\bibitem[mum()]{mumag4}
$\mu${MAG} micromagnetic modeling activity group.
\newblock URL \url{http://www.ctcms.nist.gov/~rdm/mumag.org.html}.

\bibitem[Aharoni(2000)]{aharoni2000introduction}
A.~Aharoni.
\newblock \emph{Introduction to the Theory of Ferromagnetism}, volume 109.
\newblock Clarendon Press, 2000.

\bibitem[Exl et~al.(2012)Exl, Auzinger, Bance, Gusenbauer, Reichel, and
  Schrefl]{exl2012fast}
L.~Exl, W.~Auzinger, S.~Bance, M.~Gusenbauer, F.~Reichel, and T.~Schrefl.
\newblock Fast stray field computation on tensor grids.
\newblock \emph{Journal of computational physics}, 231\penalty0 (7):\penalty0
  2840--2850, 2012.
\newblock URL \url{http://dx.doi.org/10.1016/j.jcp.2011.12.030}.

\bibitem[Exl et~al.(2014{\natexlab{b}})Exl, Abert, Mauser, Schrefl, Stimming,
  and Suess]{exl2014fft}
L.~Exl, C.~Abert, N.~J. Mauser, T.~Schrefl, H.~P. Stimming, and D.~Suess.
\newblock F{FT}-based {K}ronecker product approximation to micromagnetic
  long-range interactions.
\newblock \emph{Mathematical Models and Methods in Applied Sciences},
  24\penalty0 (09):\penalty0 1877--1901, 2014{\natexlab{b}}.
\newblock URL \url{http://dx.doi.org/10.1142/S0218202514500109}.

\bibitem[Miltat and Donahue(2007)]{miltat2007numerical}
J.~E. Miltat and M.~J. Donahue.
\newblock Numerical micromagnetics: Finite difference methods.
\newblock \emph{Handbook of magnetism and advanced magnetic materials}, 2007.

\end{thebibliography}

\end{document}